\documentclass{article}    

\title{Contribution of Blackhole Entropy Towards Acceleration Of The Universe}  
\author{A. M. Harunar Rashid\footnote{ Department of Physics, Dhaka University,Dhaka, Bangladesh}, and A. L. Choudhury \footnote{ Dept. of Chemistry and Physics, Elizabeth City State University, Elizabeth City, NC 27909, USA}\footnote{email: alchoudhury@mail.ecsu.edu}
}         

\begin{document}           

\maketitle                 
\begin{abstract}
{In a recent model Choudhury proposed that a collapsing matter generates an adiabatic pressure which can be used to explain the acceleration of the physical universe under certain approximation. In this work we superimpose the core pressure generated through the entropy of the blackhole. We have ignored the blackhole internal energy change during this process. We have then calculated both Hubble parameter and the deacceleration parameter. We have shown that the Hubble parameter is positive and the deacceleration parameter   flucuates with time, meaning for certain period the universe accelerates and decelerates for other time.}
\end{abstract}
\section{Introduction} 
{ \quad In a recent paper Choudhury[1] has shown that collapsing matter which ultimately turns into a blackhole can influence the universe surrounding it. This matter exerts a pressure following the classical adiabatic gas law on the physical world surrounding it. He has shown that under certain conditions the physical universe accelerates[2].}
{\par  In this paper we want to incorporate the idea that the collapsing matter has a blackhole core and has an entropy associated with it. The entropy is proportional to the surface area of the blackhole as has been shown by Beckenstein and Hawking[3]. We assume that this entropy satisfies the thermodynamical relation}
\begin{equation}
du = TdS-pdV
\end{equation}
{\par Assuming that the change of internal energy is negligible, we can use the quation to determine the pressure emanating from the blackhole to the surrounding space. We incorporate this pressure to calculate the Hubble parameter and deacceleration parameter.We assume[4] that the scale factor of the blackhole $\beta(t)$ is shrinking according to the rule}
\begin{equation}
\beta(t)=\beta_0 e^{-\Delta t}.
\end{equation} 
{\par This enables us to find that the Hubble parameter is positive, indicating that the physical universe is expanding. It turns out that the deacceleration parameter fluctuates. There is a time interval where this parameter is negative, implying that the physical universe is expanding in an accelerating phase. But there is also time interval when decelerates.
\par In section 2 we show how we have derived the pressure due to the blackhole entropy. In section 3 we have calculated the time dependent Hubble and deacceleration parameters. In section 4 we discussed the implication of our calculation. }

\section{ Pressures From Collapsing Matter and Entropy}
{\quad Choudhury[1] has shown that if we assume that the collapsing matter surrounding the blackhole behaves like an adiabatic gas , then pressure $P_a$ evolved can be written as }
\begin{equation}
P_a=Bcos^{-6\gamma}(\frac {{\surd\alpha} t} 2)
\end{equation}
{In the above equation $\rho^c(0)$ is the core density at time t=0, and $\gamma$ is the adiabatic gas law constant.
\par We now superimpose the contribution of the pressure from the blackhole entropy as we have indicated in the Eq.(1). Here we assume that the internal energy change dU is negligible, we get the pressure originated from the entropy is given by}
\begin{equation}
P_e=\frac {TdS} {dV}.
\end{equation}
{However we know that the blackhole entropy as shown by Beckenstein and Hawking}
\begin{equation}
dS= TKdA, 
\end{equation}
{where K is a constant A is the surface area of the event horizon.
Therefore}
\begin{equation}
P_e=TK \frac {dA} {dV},
\end{equation}
{where  V is the volumeof the blackhole. If we now recall the scale factor of the blackhole to be $\beta(t)$, we can show that}
\begin{equation}
P_e=\frac {2TK} {\beta(t)}.
\end{equation}
{We conjecture that $\beta(t)$ is shrinking according to  the Eq.(2). As a result the final form lf $P_e$ becomes }
\begin{equation}
P_e=\frac {2TKe^{\Delta t}} {\beta_0}.
\end{equation}
\section{Hubble And Deacceleration Parameters}
{\par As shown by Choudhury[1] that for a physical world with a collapsing core , the Hubble parameter turns out to be}
\begin{equation}
H(t)=\frac{\dot R} R=\surd(\gamma \rho+2\gamma(P_a +P_e)-\frac {2\sigma} {R^2})
\end{equation}
{with $P_a$ and $P_e$ are given by Eq.(3) and Eq.(8) and R(t) is the scale factor of the physical universe. We get for large time $t_0$ }

\begin{equation}
H(t_0)=\surd(\frac{\gamma \rho} 3+\frac{4\gamma TKe^{\Delta t}} {\beta(t_0)}).
\end{equation}
{This quantity is positive, therefore the universe is expanding. On the other hand Choudhury has shown that the deacceleration parameter is given by}
\begin{equation}
q_0(t)={\frac1 2}[\frac{\gamma(\rho+3(P_a +P_e)} {\gamma(2\rho+12[P_a+P_e])-\frac{2\sigma} {R^2}}].
\end{equation}
{We know that $P_a$ and $P_e$ are given by the relations  Eq.(3) and Eq.(8). The scale factor R dependent term vanishes for large t. We can thus simplify the  large $t_0$ indicating the present time. The deaccelation parameter for present time beomes}
\begin{equation}
q_(t_0)={\frac1 2}[1-\frac{6TKe^{\Delta{t_0}}+B{\beta_0}cos^{-6\gamma}({\frac{{\surd\alpha} t_0} 2})} {\rho\beta_0+(12TKe^{\Delta t_0}+B{\beta_0}cos^{-6\gamma}(\frac{\surd{\alpha} t_0} 2))}].
\end{equation}
{\par The condition for negative $q(t_0)$ is given by the relation}
\begin{equation}
\rho<-3{\frac 3 {\beta_0}}[2TKe^{\Delta{t_0}}+B{\beta_0}cos^{-6\gamma}(\frac{\surd{\alpha} t_0} 2)].
\end{equation}
{We assume in the above equation that the  quantity $\alpha$ is so small that for large $t_0$ the following condition is satisfied}
\begin{equation}
\surd\alpha{ t_0}<4\pi.
\end{equation}
{ For the range }
\begin{equation}
\pi<\surd\alpha{t_0}<3\pi,
\end{equation}
{ the condition Eq.(13 )can be expressed as}
\begin{equation}
cos(\frac{\surd{\alpha} t_0} 2)<(-1)^{ {6\gamma}}{[\frac{2TKe^{\Delta t_0}} {B\beta_0}]}^{{6\gamma}}.
\end{equation}
{\par If we take $6\gamma$ to be odd, we can easily satisfy the above relation for small T and large B. Under the above restriction we can attain the acceleration of physical universe.}
{\par But one attractive aspect of the model is the fact that the physical universe will have a braking phase. This has been claimed by some Australian observers.}
\section{Concluding Remarks}
{\quad We have shown in this paper that if we superimpose the contribution of the pressure associated with the blackhole entropy on the adiabatic pressure from the collapsing matter, which ultimately forms the blackhole. would also under certain restriction make the physical universe accelerate. We are aware of the fact that this acceleration is achieved under the restriction that $\surd{\alpha} t_0$ is small. Since $t_0$ , the present time of the physical universe is very large, the quantity $\alpha$ has to be very small.  From Eq.(9) in referenc [1] we note this can be satisfied because G is small and $\rho^c (0)$ can be considered a small quantity. We want to stress the point that the acceleration of universe can be attributed to so some physical dynamical process. It will be very interesting to see how this tally with the observational data.
\par We can also try to improve improve the computation carried out by Choudhury. Whether these results would be still valid remains to be seen. Work is in progress along this line.   }

\section{Reference}
\begin{enumerate}
\item A. L Choudhury: Influence of collapsing matter on the envelopinge expanding universe; arXiv.gr-qc/0506009v1, 1 Jun 2005.
\item N. Bachcall, J. P. Ostriker, S. Perlmutter, and P. J. Steinhardt, Science, 284, 1481 (1999).
\item R. Penrose: The Road To Reality: Alfred A. Kopf, New York;715 (2004).
\item A. M. Harunar Rashid, A. Momen, and A.L. Choudhury: An accelerating universe around a blackhole; arXiv.gr-qc/0608094v1, 20 August 2006.
\end{enumerate}          

\end{document}